\documentclass[twocolumn]{aastex6}
\slugcomment{To appear in JAA special issue on Astrosat}
\shorttitle{CPM on the $AstroSat$ mission}
\shortauthors{Rao et al}

\begin{document}


\title{Charged Particle Monitor on the AstroSat mission}    



\altaffiltext{1}{Email: \url{arrao@tifr.res.in}}
\altaffiltext{2}{Tata Institute of Fundamental Research, Mumbai, India}
\altaffiltext{3}{Inter University Centre for Astronomy \& Astrophysics, Pune, India}
\altaffiltext{4}{Nicolaus Copernicus Astronomical Center, Polish Academy of Sciences, Warsaw, Poland}
\altaffiltext{5}{Vikram Sarabhai Space Centre, Thiruvananthapuram, India}
\altaffiltext{6}{Physical Research Laboratory, Ahmedabad, India}
\altaffiltext{7}{ISRO Satellite Centre, Bengaluru, India}

\author{A. R. Rao\altaffilmark{1,2}} 
\author{M. H. Patil\altaffilmark{2}}
\author{Yash Bhargava\altaffilmark{2,3}}
\author{Rakesh Khanna\altaffilmark{2}}
\author{M. K.  Hingar\altaffilmark{2}}
\author{A. P. K. Kutty\altaffilmark{2}}
\author{J. P. Malkar\altaffilmark{2}}
\author{Rupal Basak\altaffilmark{2,4}}
\author{S. Sreekumar\altaffilmark{5}}
\author{Essy Samuel\altaffilmark{5}}
\author{P. Priya\altaffilmark{5}}
\author{P. Vinod\altaffilmark{5}}
\author{D. Bhattacharya\altaffilmark{3}}
\author{V. Bhalerao\altaffilmark{3}}
\author{S. V. Vadawale\altaffilmark{6}}
\author{N. P. S. Mithun\altaffilmark{6}}
\author{R. Pandiyan\altaffilmark{7}}
\author{K. Subbarao\altaffilmark{7}}
\author{S. Seetha\altaffilmark{7}}
\author{K. Suryanarayana Sarma\altaffilmark{7}}

\begin{abstract}
Charged Particle Monitor (CPM) on-board the  $AstroSat$ satellite is an instrument designed  to detect the flux of charged particles at the satellite location. A Cesium Iodide Thallium (CsI(Tl)) crystal is used with a Kapton window to detect protons with energies greater than 1 MeV. The ground calibration of CPM was done using gamma-rays from radioactive sources and protons from particle accelerators. Based on the ground calibration results, energy deposition above 1 MeV are accepted and particle counts are recorded. It is found that CPM counts are steady and the signal for the onset and exit of South Atlantic Anomaly (SAA) region are generated in a very reliable and stable manner.

\end{abstract}




\section{Introduction}
\label{intro}

Satellites in the Low Earth Orbit (LEO) pass through the trapped radiation belts of the South Atlantic Anomaly (SAA). In this region, the particle environment can change very drastically within a few tens of seconds. A model for the SAA can be created from data accumulated by previous missions, but that does not account for variable influences like solar flares. Previous studies indicate a significant drift rate of SAA region \citep{JGRA:JGRA13402, 4342532}.\par

The charged particles (mostly consisting  of protons) cause adverse effects like the saturation of the detectors, detector dead-time (after SAA), aging of the detectors etc. Hence, in SAA region, most of the X-ray detectors need to be switched off. To optimise the operation time of the detectors, this region needs to be monitored and the entry and exit times of the satellite in the SAA region need to be estimated accurately. \par

$AstroSat$ is a multi-wavelength observatory which was launched on 2015 September 28 in a LEO with an altitude of 650 km and an inclination of $ 6^{\circ} $. Primary instruments of  $AstroSat$ include Soft X-ray Telescope (SXT), three Large Area X-ray Proportional Counters (LAXPCs), Cadmium-Zinc-Telluride Imager (CZTI), Ultra-Violet Imaging Telescope (UVIT) and Scanning Sky Monitor (SSM) \citep{2014SPIE.ASTROSAT}. Many of these instruments are sensitive to charged particles  in the SAA region. With low source counts of X-ray sources, it is desirable that the observation time of each instrument is as high as possible. For better optimisation, a monitor for particle count rate, called Charged Particle Monitor (CPM), has been installed as an auxiliary payload in the  $AstroSat$ satellite. \par

In previous missions, to alleviate the problem of particle bombardment, particle monitors were incorporated alongside the primary instruments. Rossi X-ray timing explorer(RXTE) included a particle monitor with the High
Energy X-Ray Timing Experiment (HEXTE) \citep{1998ApJ..RXTE} and BeppoSAX mission with Phoswich Detection System (PDS) \citep{1997A&AS.BeppoSAX}. Both particle monitors used plastic scintillators coupled with a Photo-Multiplier Tube (PMT). In both cases, the particle monitor sent a signal to reduce the voltage of the PMT connected to the Phoswich detector when the satellite was in the SAA region. \par

CPM in  $AstroSat$ is designed to measure the count rate of charged particles at the satellite location. It is sensitive to protons above 1 MeV. Following sections detail the device design, electronics and calibration procedure used on ground and in flight for the  assessment of the device.

\section{Detector Design}
\label{det-des}
Solid state detectors combined with a photo-diode are used for the detection of the high energy particles. They are preferred over standard photo-multiplier tube (used in HEXTE particle monitor \citep{1998ApJ..RXTE}) for their compactness and stability in gain \citep{PhysRev.122.815}. A 10 mm cube of Cesium Iodide Thallium activated (CsI(TI)) crystal (wavelength = 550 nm) with Teflon reflective material is coupled to the same area window of Si-PIN diode. This photo-diode has a broadband response in the visible spectrum (average efficiency 50$\%$) with high sensitivity, low dark current and good energy resolution. To achieve the lower energy threshold requirement, the top side of the detector is covered with a very thin sheet of Kapton / aluminium. A 25 mil Kapton gives an effective low energy threshold of 1.2 MeV. The electronic threshold is programmable and has been kept at 0.5 MeV. CsI crystal has higher detection probability for gamma-rays \citep{PhysRev.122.815}, which has been used in the calibration of the device. \par
The typical anticipated count rate is about   1 count cm$^{-2}$  s$^{-1}$ str$^{-1}$ above 1 MeV, and it can increase by a factor of 100 to 1000   in  the  SAA region. Hence, the CPM, with an area of about 1 cm$^2$, can be sensitive to protons of typically about 1 MeV (the proton energy spectrum at these energies is quite flat and, hence, the observed count rate would be insensitive to the lower threshold, up to a couple of MeV). It is estimated that gamma-ray bursts and solar gamma-ray flares can give a measurable count rate ($>$ 10 counts/s) above 0.5 MeV in very rare cases (occurring at a rate of less than once per year). The time-scale for detecting the entry and exit of SAA is kept large enough to ignore even such rare events. \par

\subsection{Electronic Design}
\label{ed}

When a proton strikes the detector, it causes ionisation, thereby resulting in a bunch of photons, the total number of which is proportional to the  energy of the  incident proton. The photo-diode converts these photons into an electrical signal by generating a voltage proportional to the number of  photons. Since the photodiode output impedance is high and signal level is around a few millivolts, a low noise Charge Sensitive Pre-Amplifier (CSPA) is used to process the signal. The CSPA is placed just behind the photo-diode to reduce noise. The CsI(Tl) crystal, along with Si-PIN photo-diode and CSPA, are all combined into a small compact module. \par

The output of the CSPA can be connected to a Multi-Channel Analyser (MCA) which can give the required spectral information. In the flight model, the output of the CSPA is connected to a comparator. The comparator takes another input from the Low-Level Discriminator (LLD) circuit. If the signal from the CSPA is higher than the LLD signal, the comparator sends a pulse to the gating circuit. The level of LLD is programmable from ground. The event pulse generated from the comparator is passed through a free running 16-bit counter. This counter is gated at 5 s intervals, at which it sends the accumulated counts to the Processing circuit. The gating time of 5 s is also programmable from ground. \par

The count rate is transmitted through telemetry to ground station. It is also made available to onboard users (other experiments) in a serial format. This count rate is compared with a preset value, which is programmable from ground. Whenever the count rate is greater than the preset value, an output is activated. This output is deactivated whenever the count rate goes below  the preset value. To avoid false triggering, the output is activated/de-activated only after 3 successive confirmation of count rate. These SAA warning outputs are available for other experiments. The design specifications of CPM have been summarised in Table \ref{tab:design_spec}.  \par

\begin{table}[ht]
\centering
\caption{Design Specifications of CPM}
\label{tab:design_spec}
\begin{tabular}{|l|l|}
\hline
Scintillator                    & CsI(Tl)               \\ \hline
Size                            & 10 mm $\times$ 10 mm $\times$ 10 mm \\ \hline
Light Collector                 & Photodiode with CSPA  \\ \hline
Electronic low energy threshold & 0.5 MeV               \\ \hline
Window                          & 25 mil Kapton        \\ \hline
Window Threshold                & 1.2 MeV               \\ \hline
Field of View                   & 2$ \pi $ Steradians       \\ \hline
Gating time                     & 5 s (programmable)    \\ \hline
LLD                             & 1 MeV (programmable)  \\ \hline
Count rate threshold            & 3 counts/s (programmable)\\ \hline
User output                     & Serialised count rates   \\ 
                                        &  and SAA signal\\  \hline
\end{tabular}
\end{table}

\section{Ground Calibration}\label{g_cal} 

\begin{figure*}[!htpb]
\centering
\plottwo{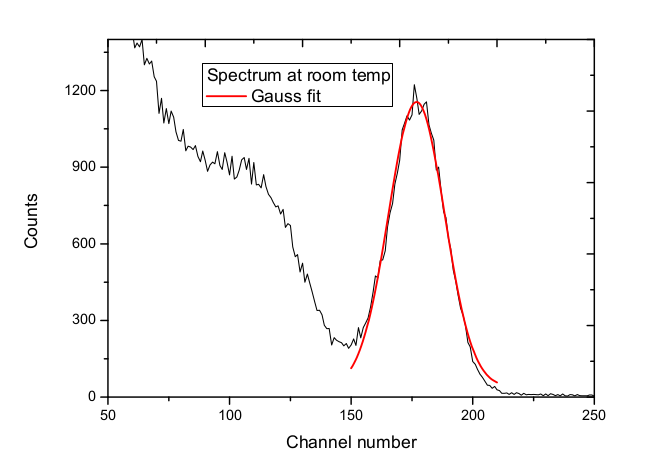}{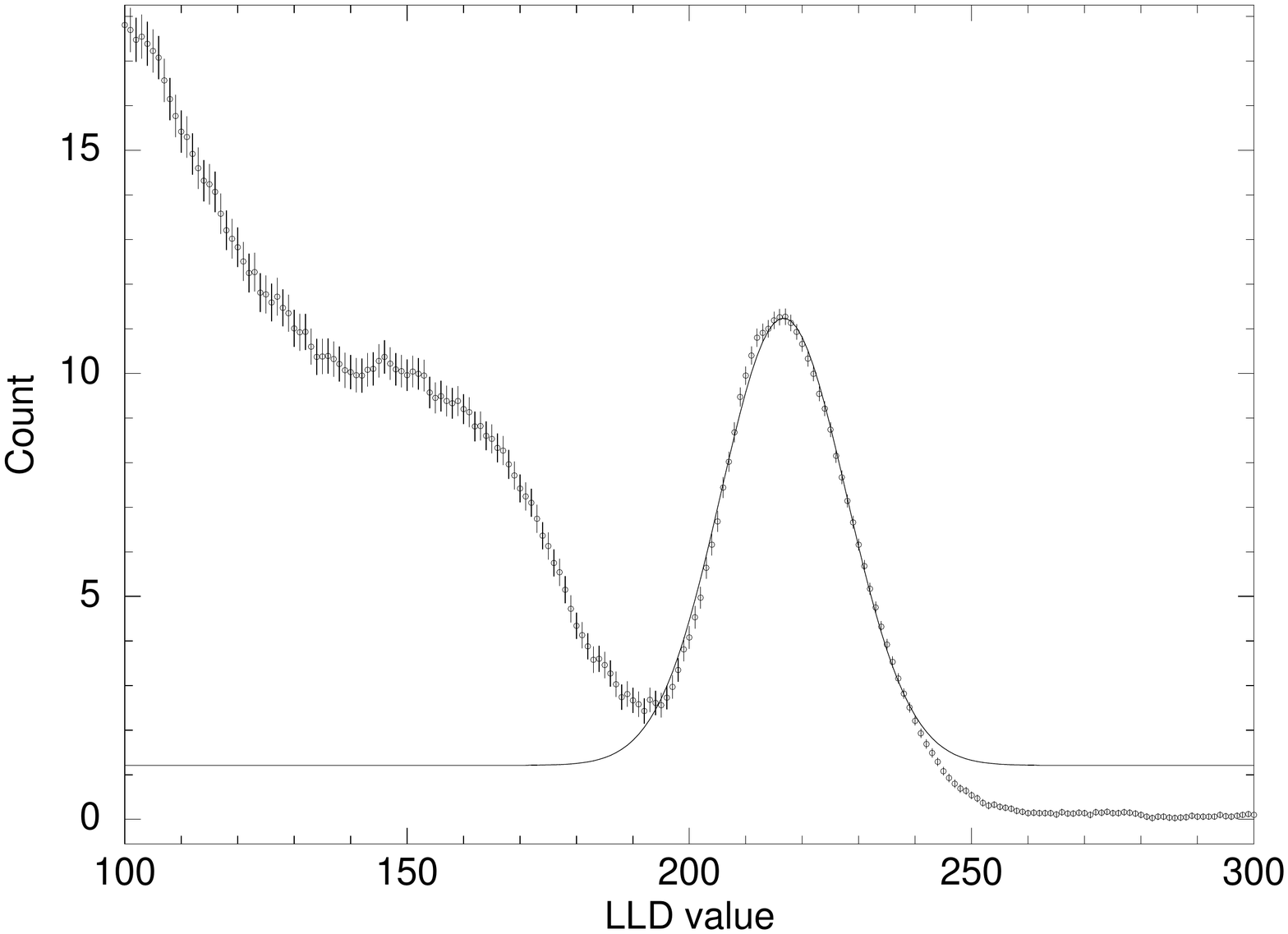}
\caption{Spectrum of radioactive source Cs$^{137}$ obtained from CPM. (a) A Multi Channel Analyser (MCA) was used to obtain the spectral information in the Engineering model (EM). (b) Spectrum for further models was obtained by changing the LLD levels and reading the counts above it. The integral spectrum thus obtained was differentiated to get the spectrum.}
\label{fig:spec} 
\end{figure*}

The calibration of the device as a suitable SAA monitor was done on ground. Spectral analysis can determine the stability of the detector as well as the response of the detector in various operational  conditions. The output of the CPM is the number of events with energy greater than the LLD. As the output of the device does not have direct energy information, different methods were used to get energy information. An MCA was used only for the Engineering model (EM). For Qualification and Flight models (QM \& FM) integral spectrum was obtained by varying the LLD, which was later differentiated to get the actual spectrum. The spectrum obtained from the MCA is shown in Figure \ref{fig:spec}\textcolor{cyan}{(a)} while the differentiated spectrum in shown in Figure \ref{fig:spec}\textcolor{cyan}{(b)}. \par
The calibration included illumination of the device with radioactive sources and determining the peak position in the differentiated spectrum by fitting a Gaussian function. The position of the peak and  the Full width at Half Maximum (FWHM) were used for the calibration of the device. To observe the response of the detector, testing was done for different sources, and to see the effect of particles on the detector, protons of different energies were bombarded on the detector. Further experiments tested the stability of the detector by observing for different time durations and in different environmental conditions. \par

\subsection{Linearity in Gamma ray response}\label{diff_src}
\begin{figure}[!ht]
\centering
\includegraphics[width = 0.4\paperwidth]{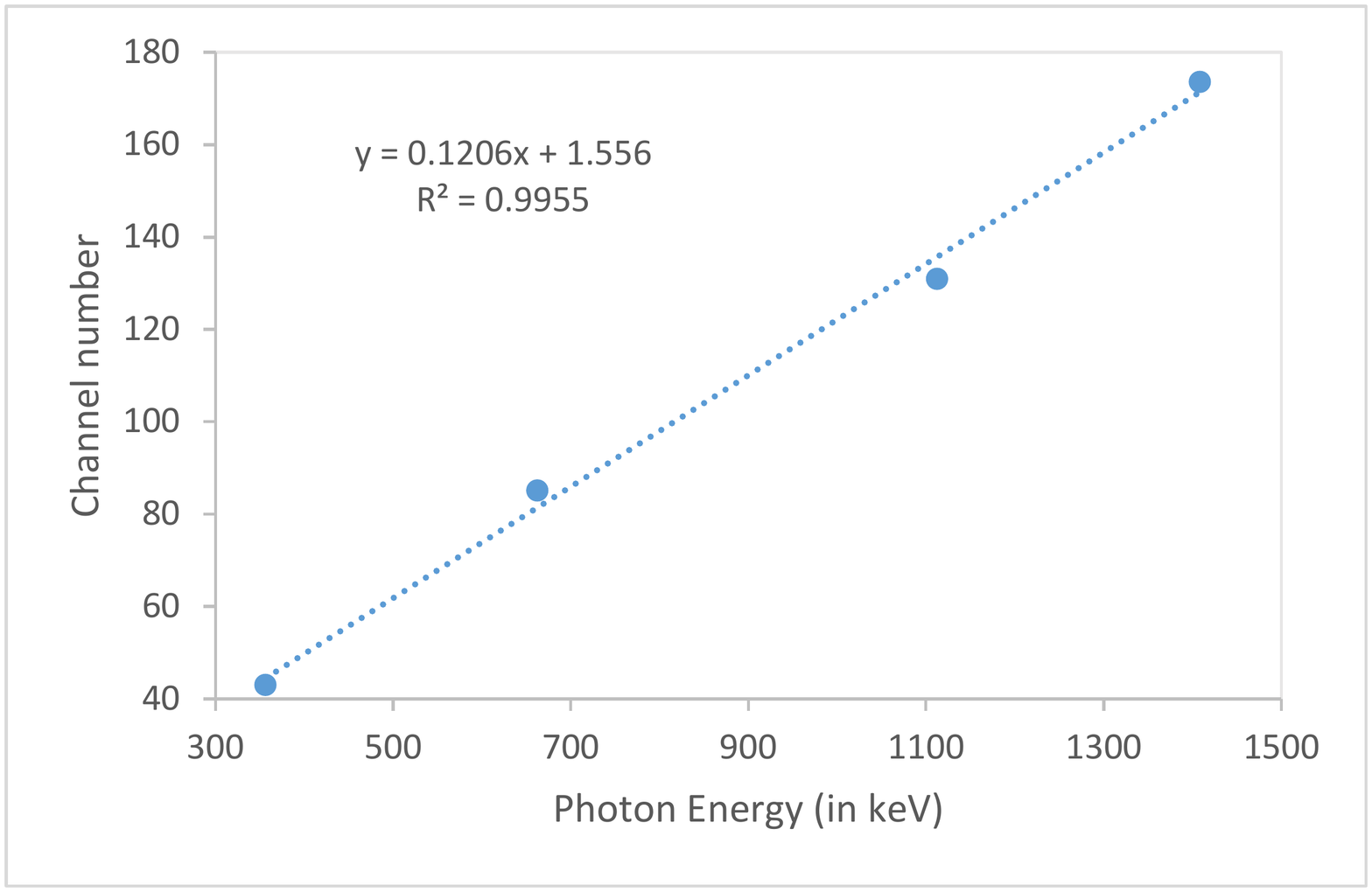}
\caption{Linearity in the peak position for different Gamma ray sources (linear regression coefficient $R^2$ of 0.9955). The detector produces photons of energy proportional to the incident radiation,
correct to about 5\%. Different photon sources were used to observe and calibrate the energy response of the detector. The energy response can be used to determine the energy of the incident protons (Section \ref{diff_src}).}
\label{fig:diff_src}
\end{figure}

The detector was tested with different radioactive sources to observe the response of the device. The gamma rays produced by these sources are close to the expected energy threshold of the detector and thus can characterize the detector performance at those energies. The sources used were Ba$ ^{133} $, Cs$ ^{137} $ and Eu$ ^{152} $. The energies of the gamma rays emitted by these sources and the corresponding peak channel observed have been tabulated in Table \ref{tab:diff_src}. The peak position followed an increasing trend and was fitted with a linear function. Figure \ref{fig:diff_src} shows the fitting function and its parameters. MCA channel 1024 corresponded to 10 V. On converting energy from the fitting, the detector response works out to around 1.20 mV/keV.

\begin{table}[ht]
\centering
\caption{Detector performance for different radioactive sources}
\label{tab:diff_src}
\begin{tabular}{|c|c|c|}
\hline
Source & Energy (in keV) & Peak Channel Number \\ \hline
Ba$ ^{133} $     & 356             & 43.1\\ \hline
Cs$ ^{137} $     & 662             & 85.2\\ \hline
Eu$ ^{152} $     & 1112            & 131.0 \\ \hline
Eu$ ^{152} $     & 1408            & 173.6\\ \hline
\end{tabular}
\end{table}

\subsection{Linearity in high-energy particle response} \label{part_sens}

Since the CPM is supposed to detect charged particles, the testing of the device was also done using  proton beams. Proton beams of energy 15, 17 and 18 MeV were generated in BARC-TIFR Pelletron facility. The proton beams for lower energy were not used as these have weaker peak energy and the data from the gamma ray calibration was used to correlate the two beams. The available proton incidence rate was around $10^{10}$ counts/s and the expected count rate in deep SAA is only $10^3$ counts/s. Thus the mounting was done off-axis to reduce the count rate. The observed energy of the beam was less due to attenuation by the air column, black paper, teflon and mylar covering of the detector. The data from the experiment is tabulated in Table \ref{tab:part_sens}. \par 

\begin{table}[ht]
\centering
\caption{Response to protons}
\label{tab:part_sens}
\begin{tabular}{|p{0.12\textwidth}|p{0.12\textwidth}|p{0.12\textwidth}|}
\hline
Energy (MeV) & Peak channel number & Error in energy (MeV) \\ \hline
6.1         & 177.5               & 0.7                  \\ \hline
10.1         & 378.0               & 0.4                 \\ \hline
11.7         & 448.7               & 0.3                 \\ \hline
\end{tabular}
\end{table}

\begin{figure}
\centering
\includegraphics[width = 0.4\paperwidth]{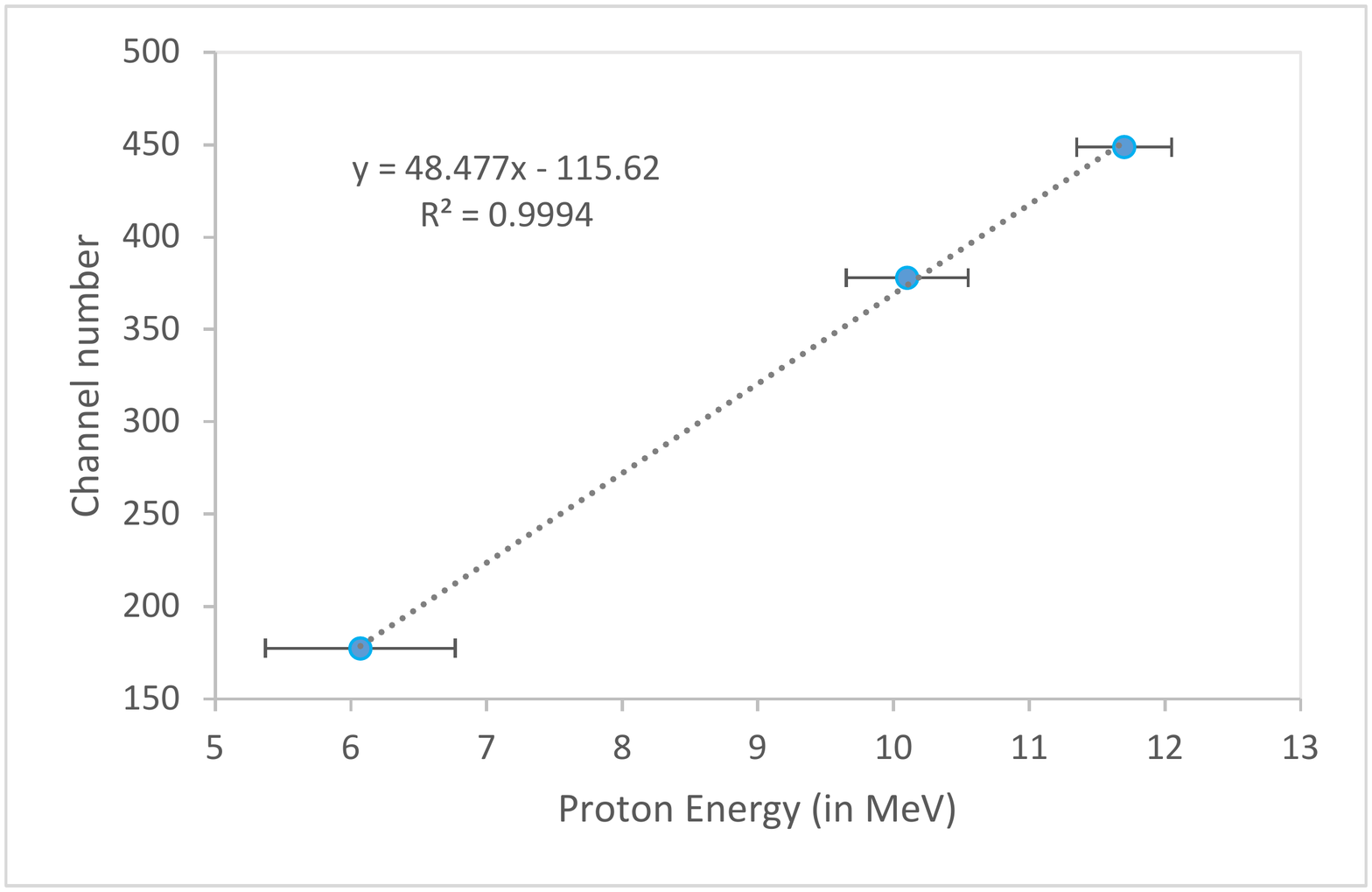} 
\caption{Linearity in the peak position for proton beams (linear regression coefficient $R^2$ of 0.9994). The photon generated in the detector on incidence of a proton is proportional to the proton's energy. Proton beams of energy 15, 17 and 18 MeV are used. Using the response derived in Section \ref{diff_src}, energy of the attenuated proton is estimated and response of proton is generated. }
\label{fig:part_sens}
\end{figure}

In the MCA, 1024 channel corresponds to 10 V. From the slope of the linear fit (48.477), the conversion rate is about 0.47 mV/keV. This value is lower than photon response as the light output of CsI (Tl) is lower for the protons as compared to photons \citep{PhysRev.122.815}. This in turn reduces the pulse height which leads to reduction in energy to channel conversion rate. 

\subsection{Variation with time}
\label{time_var}

The detector was kept in the presence of a  source for   varying durations of time to test the effect of long exposure of radiation on the crystal and it was observed that the peak position was consistent across the readings. The data from the experiment is reported in Table \ref{tab:time_var}. As can be seen from the table, for a day's observation, the values do not vary much. The difference in the two sets of observation
(taken on 2010 Oct 25 and 2010 Nov 30, respectively) can be accounted for by the absence of copper cover in the  first case. 

\begin{table*}[!ht]
\centering
\caption{Variation of peak position with time}
\label{tab:time_var}
\begin{tabular}{|c|c|c|c|c|}
              \hline
              & \multicolumn{2}{c|}{Data Set 1} & \multicolumn{2}{c|}{Data Set 2 } \\ \hline 
Real time (s) & Peak channel             & FWHM              & Peak Channel              & FWHM             \\ \hline \hline
10            & 170                      & -                 & 184                   & -                \\ \hline
100           & 168                      & 24            & 188                   & 27             \\ \hline
500           & 171                      & 31             & 187                    & 34             \\ \hline
1000          & 169                      & 32             & 186                    & 35             \\ \hline
5000          & 168                      & 33             & 187                    & 34             \\ \hline
10000         & 168                      & 33             & 187                    & 34             \\ \hline
20000         & 167                      & 32             & 188                    & 35   
\\ \hline
\end{tabular}
\end{table*}

\subsection{Variation with environment} \label{env_var}

\begin{figure*} [ht]
\centering
\includegraphics[height=0.5\paperheight]{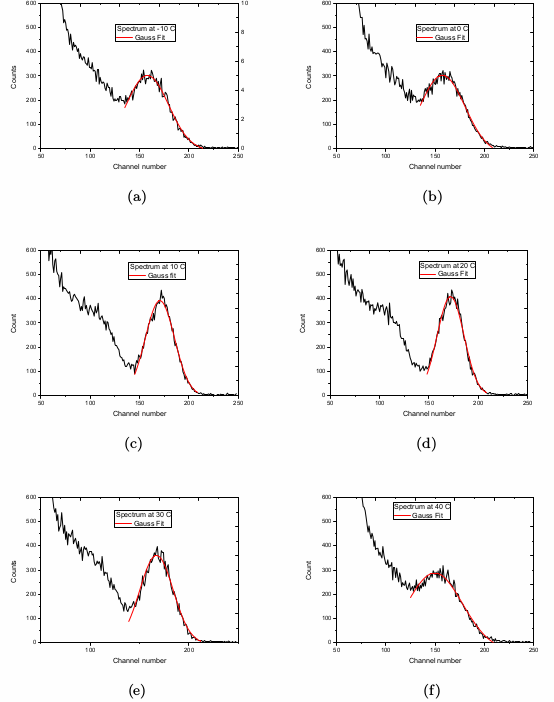}
\caption{Spectrum for the temperature range -10$ ^{\circ} $C to 40$ ^{\circ} $C with interval of 10$ ^{\circ} $C. The detector experiences large range of temperature during onboard operation. To simulate that the detector was tested in Thermo-Vacuum chamber to check the stability of the system and observe the variation in the response of the detector.}
\label{fig:env_var}
\end{figure*}

Testing at  different environmental conditions  was done in a Thermo-Vacuum chamber. The temperature was varied from -10$ ^{\circ} $C to 40$ ^{\circ} $C. The data from the experiment is tabulated in Table \ref{tab:env_var}. The detector shows degradation at   the extreme conditions, but is stable at the ambient conditions. The spectrum obtained for the same is shown in Figure \ref{fig:env_var}. Such a behaviour of the detector is expected as the detector was of commercial grade. 

\begin{table*}[!ht]
\centering
\caption{Variation of Peak position with Environment}
\label{tab:env_var}
\begin{tabular}{|c|c|c|c|c|c|}
\hline
Environment & Count rate (s$ ^{-1} $) & Peak position & FWHM  & Peak height & Sensitivity \\ \hline \hline 
-10$ ^{\circ} $C     & 894             & 158     & 51 & 312      & 8            \\ \hline
0$ ^{\circ} $C       & 961            & 157       & 54 & 327      & 7           \\ \hline
10$ ^{\circ} $C      & 330            & 171        & 35 & 394      & 22           \\ \hline
20$ ^{\circ} $C      & 311             & 172        & 32 & 408      & 23          \\ \hline
30$ ^{\circ} $C      & 400            & 167        & 40 & 366      & 18         \\ \hline
40$ ^{\circ} $C      & 1386         & 150       & 67 & 323    & 5           \\ \hline
R.T.(16$ ^{\circ} $C) & 288             & 177        & 27 & 1117     & 24          \\ \hline
\end{tabular}
\end{table*}

\section{Onboard testing}\label{onboard}

\begin{figure*}[!ht]
\centering
\includegraphics[width=0.6\paperwidth]{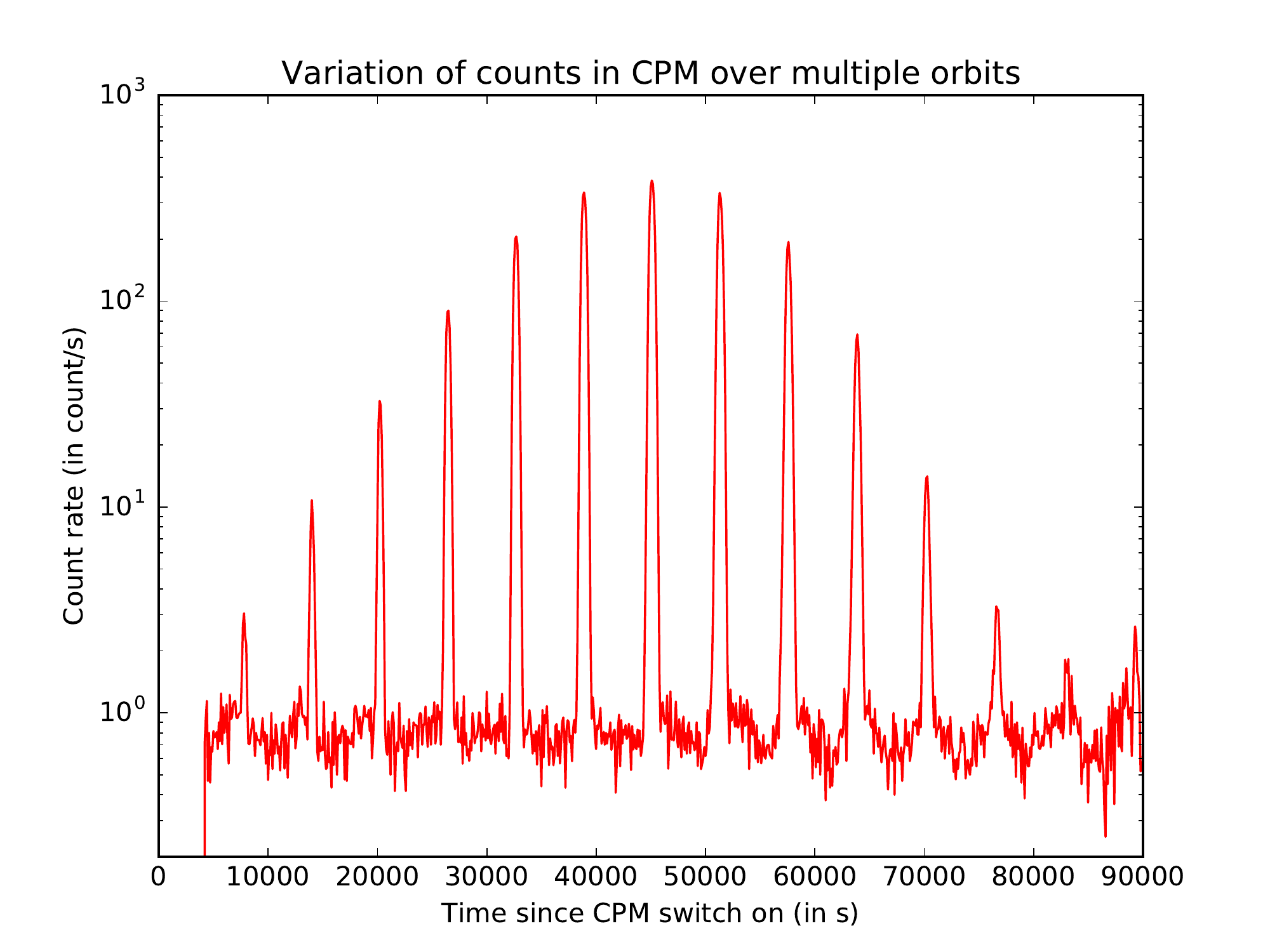}
\caption{Light curve of count as observed by CPM. In the light curve the recurring peaks denote the section of SAA region traversed in an orbit. The highest peak denotes when the satellite ventured deepest into the SAA region. Varying peak height is caused due to the  the orbital variation of satellite. }
\label{fig:lc}
\end{figure*}

The CPM started its operation on 29 September 2015 during the satellite's 19$^{th}$ orbit. The count rates from the CPM were available immediately through telemetry. The count rates for most part of the orbit varied from 0.4 to 1 count/s. The orbit of the satellite is inclined at 6 degrees, due to which the satellite periodically passes through the SAA region. Since each satellite orbit traces a different ground path,  the duration of the time spent in SAA region varies across the day. The region of SAA through which the satellite passes is non uniform in nature and thus, over the course of a day, the counts in CPM vary as shown in Figure \ref{fig:lc}. The maximum counts detected by CPM is $ \sim $400 counts/s, which occurs when the satellite passes through deep SAA region. \par

\par

\begin{figure*}[!ht]
\centering
\includegraphics[width=0.6\paperwidth]{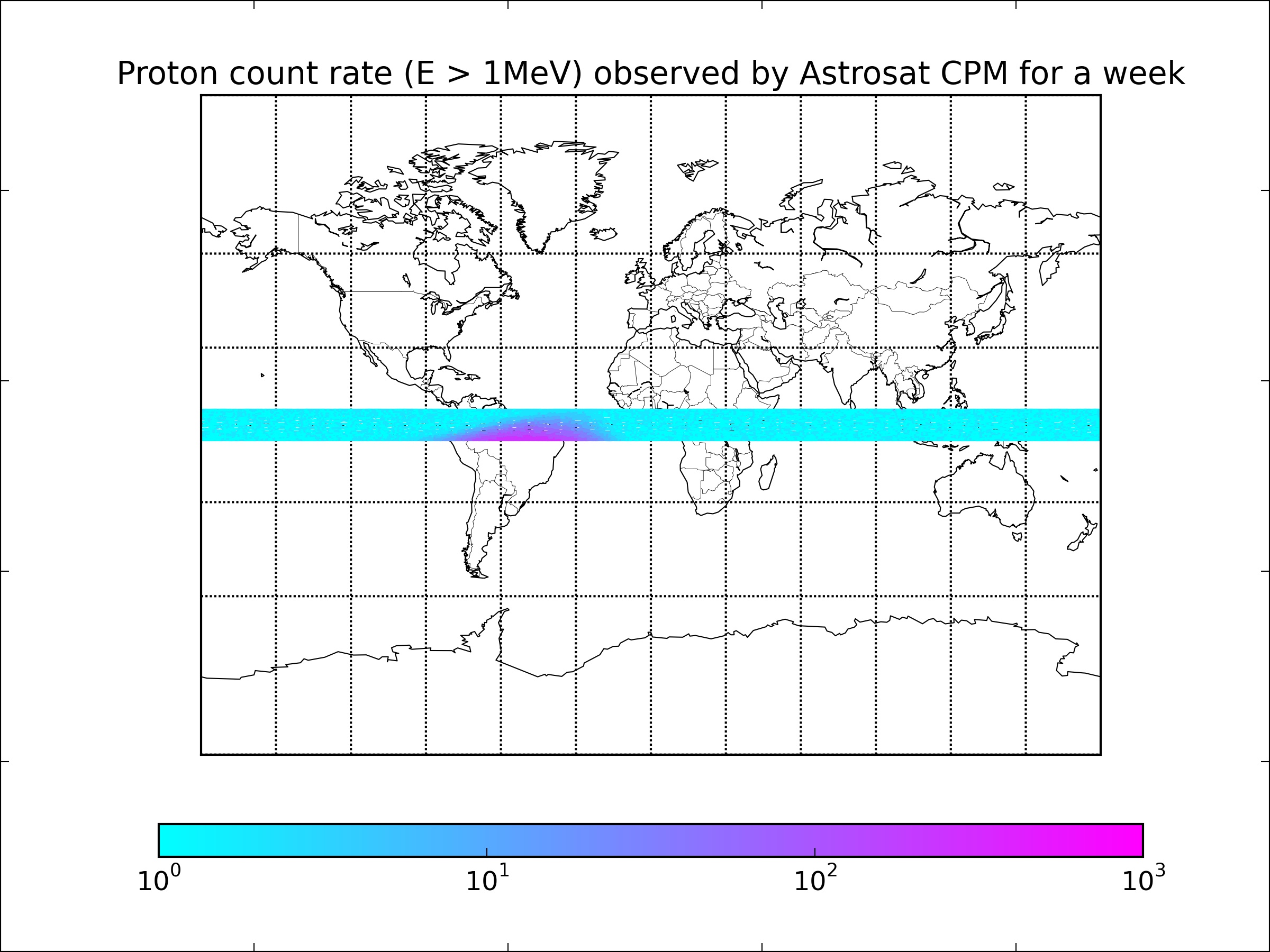}
\caption{Plot of CPM counts with latitude and longitude on World map. Using the time stamp of CPM counts, the satellite position over the earth was predicted and plotted over the map. The higher counts are concentrated over the SAA region and the low counts are smoothly distributed across the non-SAA region.}
\label{fig:plot}
\end{figure*}

The primary purpose of the CPM is to detect the regions of SAA and send a signal to the devices onboard  $AstroSat$ to switch off their respective HV. To get an idea of the shape of the SAA region at the altitude of  $AstroSat$, the counts were plotted with the latitude and longitude of the satellite. Due to the orbital variation, the region of SAA that the satellite passes through varies across days as well. Figure \ref{fig:plot} shows the variation of the counts with the position above Earth. The figure was compiled using the data from about 100 orbits ($\sim $7 days). \par

\par

\begin{figure*}[!ht]
\centering
\includegraphics[width=0.55\paperwidth]{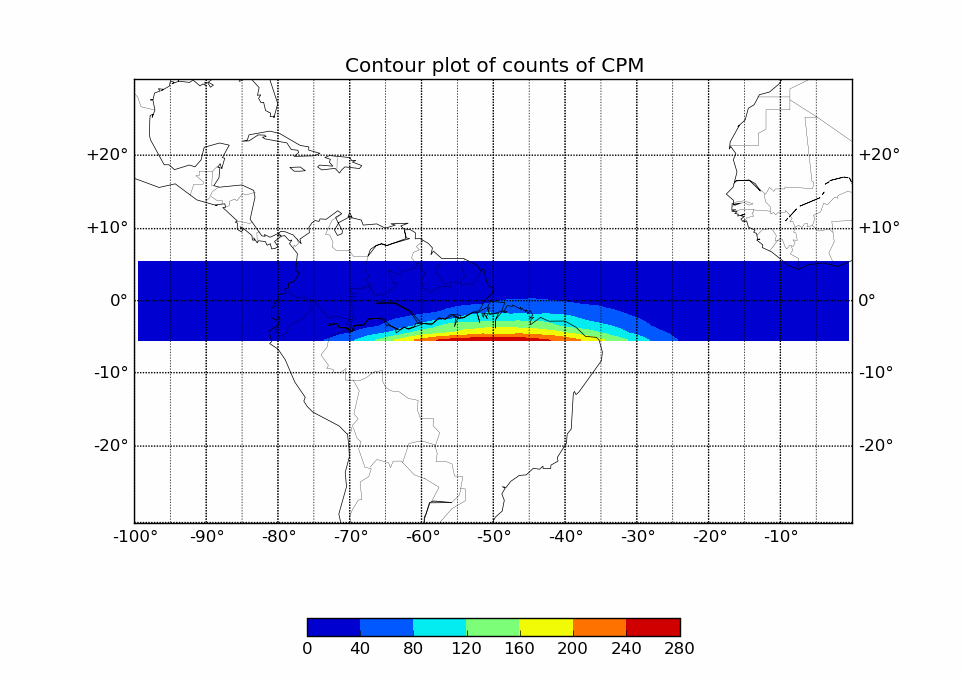}
\label{fig:contour}

\includegraphics[width=0.6\paperwidth]{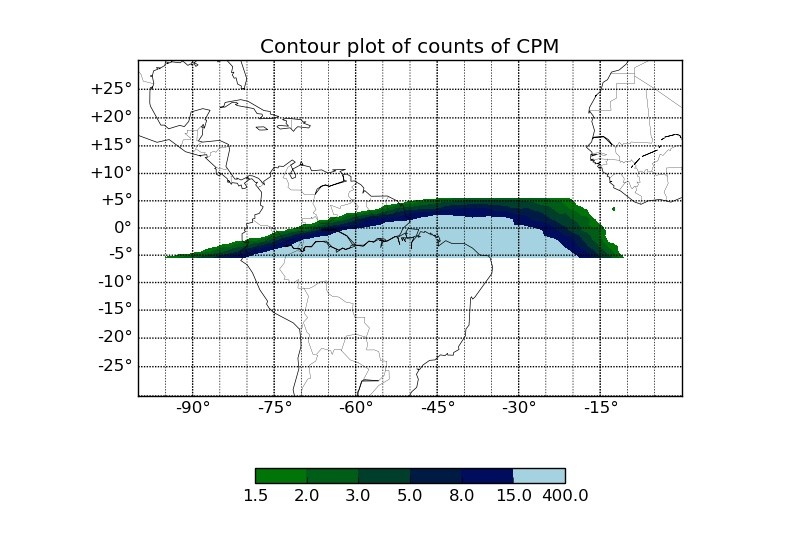}
\caption{Contour maps of SAA region.  CPM count rate data for a month  was spatially binned and contour maps were generated. The contour maps show variation of count rates from $\sim1 $ in non-SAA region to 280 counts/s on an average in figure (a). The boundary for SAA used in  $AstroSat$ is 3 counts/s which is not visible in full range of counts. For boundary selection smaller range of counts was selected (shown in figure (b)).}
\end{figure*}

As more data was colleccted, the gaps in the satellite position could be filled and contour maps of the SAA region were created. Figure \ref{fig:contour}\textcolor{cyan}{(a)} shows the contour map for the month of November 2015. The threshold kept for  $AstroSat$ instruments is 3 counts/s, which cannot be distinguished in the figure. Hence, a more detailed plot of the SAA boundary is depicted in Figure \ref{fig:contour}\textcolor{cyan}{(b)}. \par

\begin{figure*}[!ht]
\centering
\includegraphics[width=0.55\paperwidth]{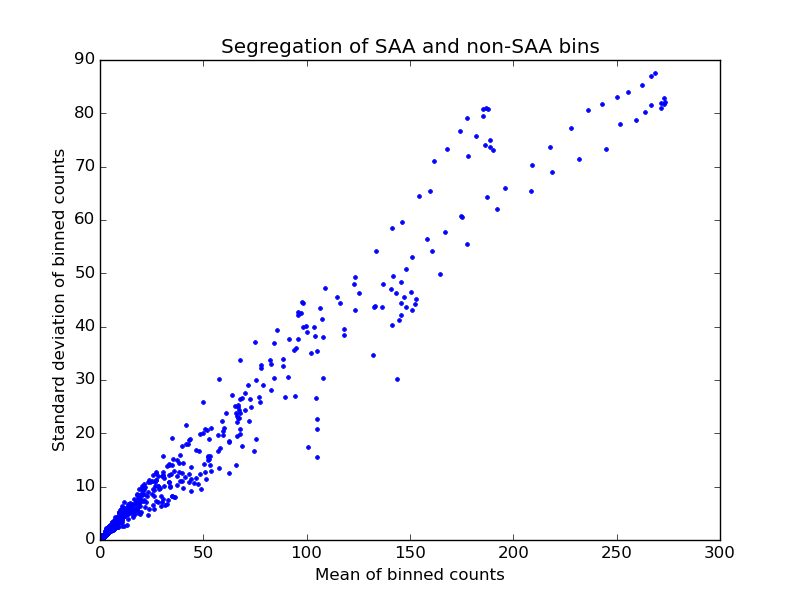}
\label{fig:stat}

\includegraphics[width=0.55\paperwidth]{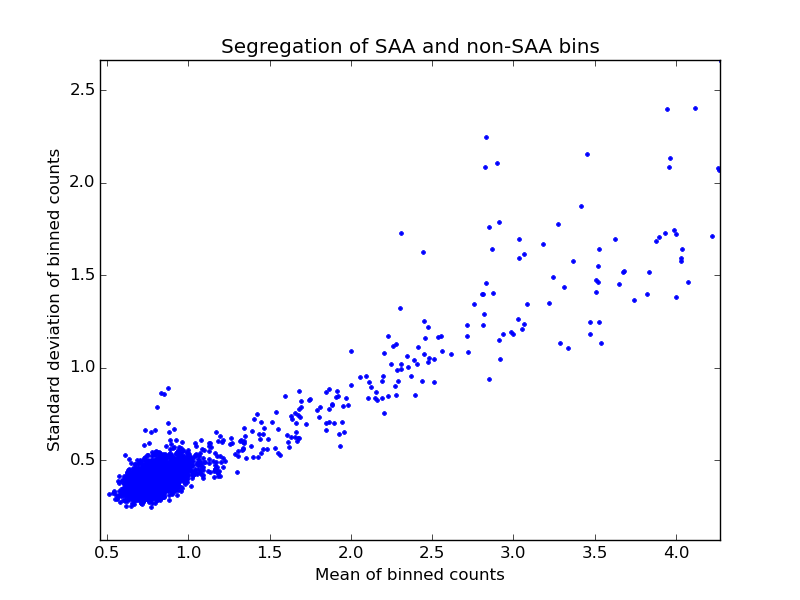}
\caption{Mean vs Standard deviation of spatially binned counts of CPM. Figure (a) shows the complete plot; Figure (b) shows the plot near the origin. For CPM the non-SAA region has low count rates ($\stackrel{<}{_\sim}1$ count/s). For most of the counts near 1 the standard deviation is also low indicating a stable behaviour in non-SAA region.}
\end{figure*}

During flight, the detector is continuously bombarded with particles which can cause stability issues in the detector. To evaluate the stability, the flight range was divided into square degree bins. The Mean and the standard deviation of the count rate were evaluated for each bin. For non-SAA region, the mean count rate is expected to be less than 1~s$^{-1}$. For SAA region, higher counts, and thus higher mean is expected. A plot of the Mean and the Standard deviation is shown in Figure \ref{fig:stat}\textcolor{cyan}{(a)} while the non-SAA region is highlighted in Figure \ref{fig:stat}\textcolor{cyan}{(b)}. The points of non-SAA region show a clumping near the origin in Figure \ref{fig:stat}\textcolor{cyan}{(b)}, which is indicative of the smoothness of the non-SAA region.\par

\section{Conclusions}
The ground calibration successfully established the stability of CPM and the onboard testing has demonstrated the capability of CPM as a monitor of the SAA region. The CPM can successfully demarcate the SAA boundary based on the primary instrument requirement. Based on the CPM data, a hard boundary for the SAA region can be used in case CPM trigger fails to alert the instruments.

 \par




\bibliographystyle{apj}

\bibliography{ref}

\end{document}